# Controlling the Coupling Strengths in Nanophotonic Networks using Modified Yagi-Uda Nanoplasmonic Antennas

Vidar flodgren[1], Anders Mikkelsen[1*]

[1] *Department of Physics and NanoLund, Lund University, 22100 Lund, Sweden*

* *Corresponding author* anders.mikkelsen@fysik.lu.se

**ABSTRACT:**

On-chip communication in optical neural networks is commonly done via waveguides which results in large system footprints. A compact alternative is to broadcast light signals between nano-optoelectronic components in free space and tailor the light field distribution using sub-wavelength nanophotonics. We propose and simulate nanoplasmonic metal structures in combination with III-V nanowire emitters and receivers to create an optical network in which the shape and position of these nanostructures create varying weights between the nano-optoelectronic nodes. Using Finite Difference Time Domain modelling, we investigate systems of experimentally verified nanowire optoelectronics combined with nanoplasmonic structures that can be made in the same fabrication step as electrical contacts powering the nanowires. We investigate both individual nanowire/antenna devices as well as networks corresponding to two layers in a neural network. We show that directed communication from a nanowire node can be significantly altered using Yagi-Uda antennas. Modifying the antenna with an asymmetric director component enables the directing of light several different angular directions simultaneously. We find that highly variable complex weight distribution between the connections in two seven node layers can be achieved depending on the combined geometry of the antenna components. The possible weight distributions in compact layers of nanowire nodes could be used for creating a variety of neural networks with complex connectivity. The concepts can be generalized to other types of nanoscale emitter/receiver systems.

## Introduction

Realizing artificial neural networks using photonics is an area of significant current interest due to the promise of high speeds and low energy consumption[1–6]. Considerable work has been done using established Si on Insulator (SOI) and InP platforms in which waveguides and microscale optics can be integrated[4,6]. These show the potential for neuromorphic computing using light, however optical waveguides have large geometric dimensions compared to conventional Si chip technology.[7–10] To reduce the footprint of optical circuits and increase the efficiency in communication, an interesting strategy is to directly broadcast between neural nodes using the local design of each node instead of dielectric channels. This can be likened to the broadcast concepts used in mobile networks in comparison to classic wire-based communication. One way to achieve such directivity is the use of nanoplasmonic metal structures that focus the light in both near and far field regimes. Further, to reach the promised low energy consumption of optical solutions, nanoscale components for emitter and receiver nodes are likely necessary[1,3,4], here direct band-gap III-V nanowires integrated with Si is a promising option.

Yagi-Uda radio antennas, invented in the early 20$^{th}$ century, relies on a single active signal transmitting/detecting element, combined with a reflector and several directors to guide both inbound and outbound signals. It has much more recently been explored in nanoplasmonic systems for optical signals with promising results (using an external light in-coupling)[11]. It has also shown significant promise for on-chip optical communication[11–14]. For combination with nano-optoelectronic components, it is of particular interest as it can be introduced at the same fabrication step as any metal contacts that are needed to power the optoelectronics. Thus, it can become a natural part of the device fabrication. Still the exploration of such components together with semiconductor nanocomponents that are used for (non-linear) evaluation of optical signals is lacking.

III-V nanowires have shown promise for uses in a wide variety of applications in electronics[15–18], photonics[19–22] and quantum technology[23–25]. III-V nanowire-based components can be integrated on Si chips and function as efficient photon-electron-photon converters in an emitter receiver scheme for neural networks[26–32]. They have been proposed as components for a nanoscale neuron that can receive both inhibiting and exciting optical signals to perform a non-linear (sigmoid) evaluation and emit light[31,32]. Fast and small neural circuits can be constructed in this way which also has the promise of low energy consumption. Recently[1,3,4] communication between InP photodiode nanowires was demonstrated on Silicon[33] indicating that efficient communication across hundred micron is possible even for single nanowires. These can be combined into a complete artificial nanophotonic neuron receiving and analyzing both inhibiting and exciting optical signals[34]

In the present work we model Yagi-Uda type nanoplasmonic antennas in which the active emitter/receiver is based on III-V nanowires. Both standard Yagi-Uda configurations as well as the introduction of an asymmetric director element at the end of the antenna are investigated. The asymmetric directors of the antenna can be used to both direct the light forward as well as split the beam in several directions. We explore how different weights of the light signals from nanowires in one layer can be attributed to different nanowires in the second layer corresponding to two layers of an artificial neural network. We find that significant improvement in emitter-receiver transmission can be achieved for 1-1 communication, but also a broad range of weight between several nodes is possible.

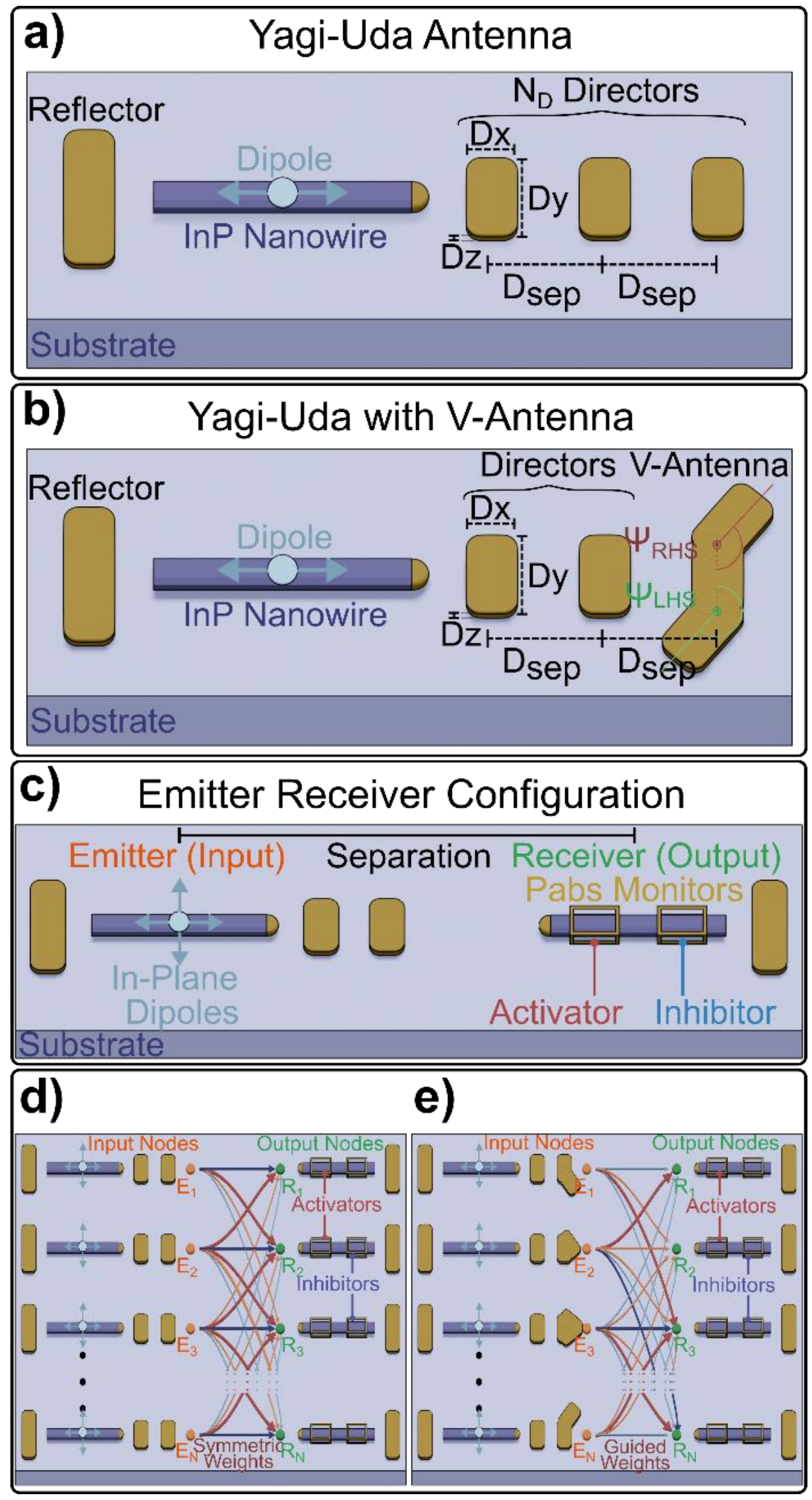

**Figure 1** - a) Antenna design, showing the metal reflector and director components and emitter nanowire of the Yagi-Uda. b) Modified Yagi-Uda design includes an end-component termed as a V-Antenna. c) System of an emitter and a receiver antenna. Dipole emitters simulate emission from the nanowire. The Power absorption monitors (Pabs) for the receiver used to estimate weights of the system. The Activator and Inhibitor regions of the nanowire indicate exciting and inhibiting regions of the nanowire receiver in an artificial neural node configuration that can be designed by doping the nanowire as described in ref. 31. d) Network with Yagi-Uda antenna that gives a symmetric emission between components. e) Network with V-Antennas that can be used to create more complex asymmetric emission patterns and thus network weights.

## Methods

Modelling was performed using Finite Difference Time Domain (FDTD) simulations in the commercial Ansys (previously Lumerical) software[35] which solves time-dependent partial differential Maxwell equations in 3D to estimate light interactions over a range of wavelengths in given nanoscale structures. These simulations allow for accurate estimates of light transmission and absorption through various device architectures. Many geometries could be simulated by automatically varying parameters such as nanowire radius and length, and oxide thickness using additional custom programming by the authors. Nanowires were modelled as cylinders with a spherical seed particle of the same radius placed at the center of the end face of the cylinder. The individual InP nanowires had a length of 3 µm, and a radius of 88 nm. The dimensions are based on highly efficient photodiodes from experimental studies[21,33,36–38].

Due to the sparse distribution of on-chip components in each simulation, an automatically designed non-conformal grid[39] is used together with individual 2-10 nm resolution mesh objects sized to tightly encapsulate each NW and metal elements. All simulations used optical functions for Au, InP, $SiO_2$ and $Al_2O_3$ from Palik[40]. All boundaries were chosen to be perfectly matched layers (PML), and careful attention was given to ensure that no Au components overlap this FDTD boundary. The underlying substrate composition also remained constant between all simulations, with 1000 nm of $SiO_2$ stacked on top of Si, which is stretched to infinity by the bottom of the PML boundary placed at its border, which also extends all other elements in the x-y plane. Convergence testing was performed until simulations stabilize at all boundary values.

For the optical simulations, a dipole source with a Gaussian wavelength emission range of 750–1050 nm was chosen to represent the measured emission spectra from a grown array of InP nanowires. [33,36] Specific regions along the receiver nanowire were defined as activator and inhibitor acting as regions that would lead to either excitation or inhibition of a nanowire based artificial neuron that output optical signals from an emitter wire[31,32]

**Results and discussion**

In Fig 1 we introduce the basic concept of our nano antenna system. The active element of the Yagi-Uda [12–14,41] consists of a 3 micron long and 100nm wide InP nanowire. The length and diameter and materials were chosen based on recent work in which these nanowires have been shown to act as efficient photodiodes[33,36]. These InP nanowires were also used as basic building blocks in artificial nanophotonic neurons[31,32,34]. By creating p-i-n photodiodes along the wires, specific regions in the wire would act to collect inhibiting and exciting optical signals which could be non-linearly evaluated by a sigmoid function and then re-emitted using another nanowire. Thus, the nanowires in the present studies represents the building blocks of such artificial neurons.

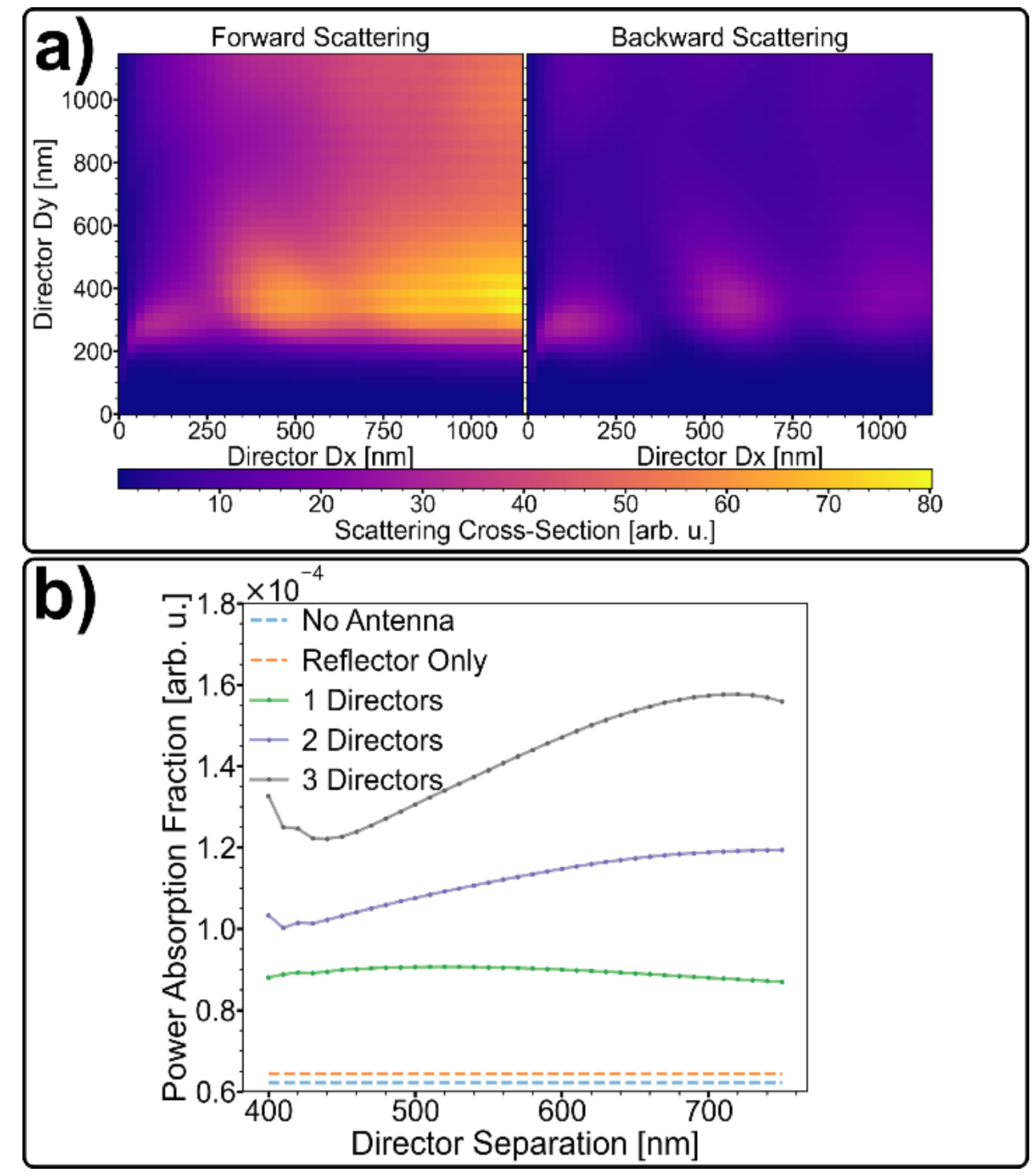

Figure 2 - a) Forward and backward scattering from a symmetric director element as in Fig 1a when varying Dx and Dy as noted in Fig 1a. Color intensity of the plot shows the forward and backward power scattering cross-section from a source with wavelength from 750-1050nm (InP nanowire emission). b) Power Absorption fraction in the receiver nanowire shown in Fig 1b from the emitter in Fig 1b. The introduction of directors increases the Power absorption by a factor of 1.5, 1.9 and 2.6 respectively for 1,2 and 3 directors.

The middle of the nanowire contains the emitting part, modelled by dipole emission from dipoles parallel and perpendicular to the nanowire. The nanowire has a gold seed particle at one end[42–45] that was included in the simulation (see SI). An Au metal reflector is placed at one end of the nanowire and 1 to 3 directors are placed at the other end, as seen inf Fig1a. We simulated symmetrical as well as modified Yagi-Uda antennas with an asymmetric third director that we name as the V-element[46,47] as seen in Fig 1b. Subsequently, a receiver nanowire with a back reflector is added to achieve an emitter-receiver system as seen in Fig 1c. Finally, we simulate a larger network of 7 emitters and 7 receivers which can represent two arbitrary layers in a neural network (seen in Fig 1d, e). In the final simulations of the full system, we added a homogeneous $Al_2O_3$ oxide layer of 230 nm thickness which acts as a quasi 2D waveguide. This significantly reduces the amount of light that escapes out of the substrate (chip) plane increasing effective transmission[33].

Starting with the symmetric Yagi-Uda antenna in Fig 1a, the ideal parameters for light guiding were derived based on a large set of simulations as seen in Fig 2. Initially we optimized the directional scattering properties of the individual director elements. We used a monitor setup described in SI Fig S1 from which we can derive the forward and backward scattering. The analysis is numerical but will have a resemblance to the analytic solutions found in Mie scattering theory for uniform particles[48–50]. We performed a wide range of simulations measuring the forward/backward scattering of the standard Yagi-Uda director component. For similarity to on-chip components, director and reflector morphology is approximated by nanostructures as found from a standard lithographic patterning[33,34]. Nano-directors are thus rounded cuboids with flattened bases, emulating the morphology that metal evaporation produces. Fig. 2a shows the total forward scattering when $D_z$=60 nm, and varying $D_x, D_y \in [10,1225] \in [10,1225]$ nm (see Fig 1a for definitions). Simulations which evaluate the effect of $D_z$ can be found in Fig. S2. As can be seen in Fig 2a, for significant forward or backward directional scattering to occur the width (Dy) of the director must be above ~200nm. In contrast, directional scattering is observed even for very narrow directors (small Dx), however we do observe a significant increase for broader directors (larger Dx). The oscillation with varying Dx can be attributed to the interplay with the wavelength range of the light source (based on InP emission). While a maximum at Dx values larger than 1000nm can be found we limited the Dx to reduce the dimension of the complete director system to the size of the typical nanowires. As for the height (Dz) of the director elements, Fig S2 shows that above a height of 40nm a good forward scattering response is observed. The maximum is found at Dz=60nm, however at higher thickness the response remains strong and only declines slowly indicating that a height between 60-100nm would work which is easily consistent with typical electrical contact heights and precision.  Taking the constraints in size into account, we can choose a nano-director shape and size based on the largest forward scattering cross-section, at $[D_x, D_y, D_z] = [485{,}350{,}60]$ nm.

Subsequently both the number of directors and their separation were varied. To evaluate the optimum number of directors and their optimal separation, a larger scale complete system of a receiver and emitter, as seen in Fig 1c, was introduced. The ideal director separation was optimized by determining the maximum sum of power absorption from the activator and inhibitor power monitors on the receiver nanowire as shown in Fig 1c. First, it can be observed that generally adding directors significantly increases the amount of light absorbed in the receiver. The highest increase (of a factor of ~2.6) is found when using three directors. A broad range of director separations will give this level of improvement, but a maximum can be found at a separation of 750nm between the centers of the director elements. It indicates that for active emitter components with a spectral profile such as such as InP nanowires, the precision in the placement of the director elements of ~100nm will be sufficient to observe the directing effect, indicating modest fabrication demands. Finally, the nanowire itself emits light directionally forward and backward in a dipole like fashion (due to its waveguiding shape), but the focusing is considerably less than with the Yagi-Uda antenna as seen in Fig 2b.

Having established the Yagi-Uda antennas ability to focus the light along the forward direction towards a receiver in a symmetric fashion (as the Yagi-Uda is symmetric), we turn to the modified Yagi-Uda. As seen in Fig 1b this antenna will have two symmetric director elements followed by the "V-antenna" element which introduce asymmetric focusing of the light. The asymmetric V-element consists of three units, a body (like the standard Yagi-Uda directors) combined with two wing elements (on each side) with dimensions as the middle element, as seen in Fig 1b. The angular orientation of the two wing elements ($Y_{LHS}$ and $Y_{RHS}$) is then varied in the simulations to create a range of light scattering patterns. In the simulation, if two of the elements overlap, all the area covered by the two elements will become the final combined element. This also means that for $\Psi_{RHS}$ or $\Psi_{LHS}$ equal to zero this corresponds to the wing not being present (as it will overlap with the middle wing). Finally, it can be noted that the combined shapes of these three elements are well suited for lithographic fabrication.

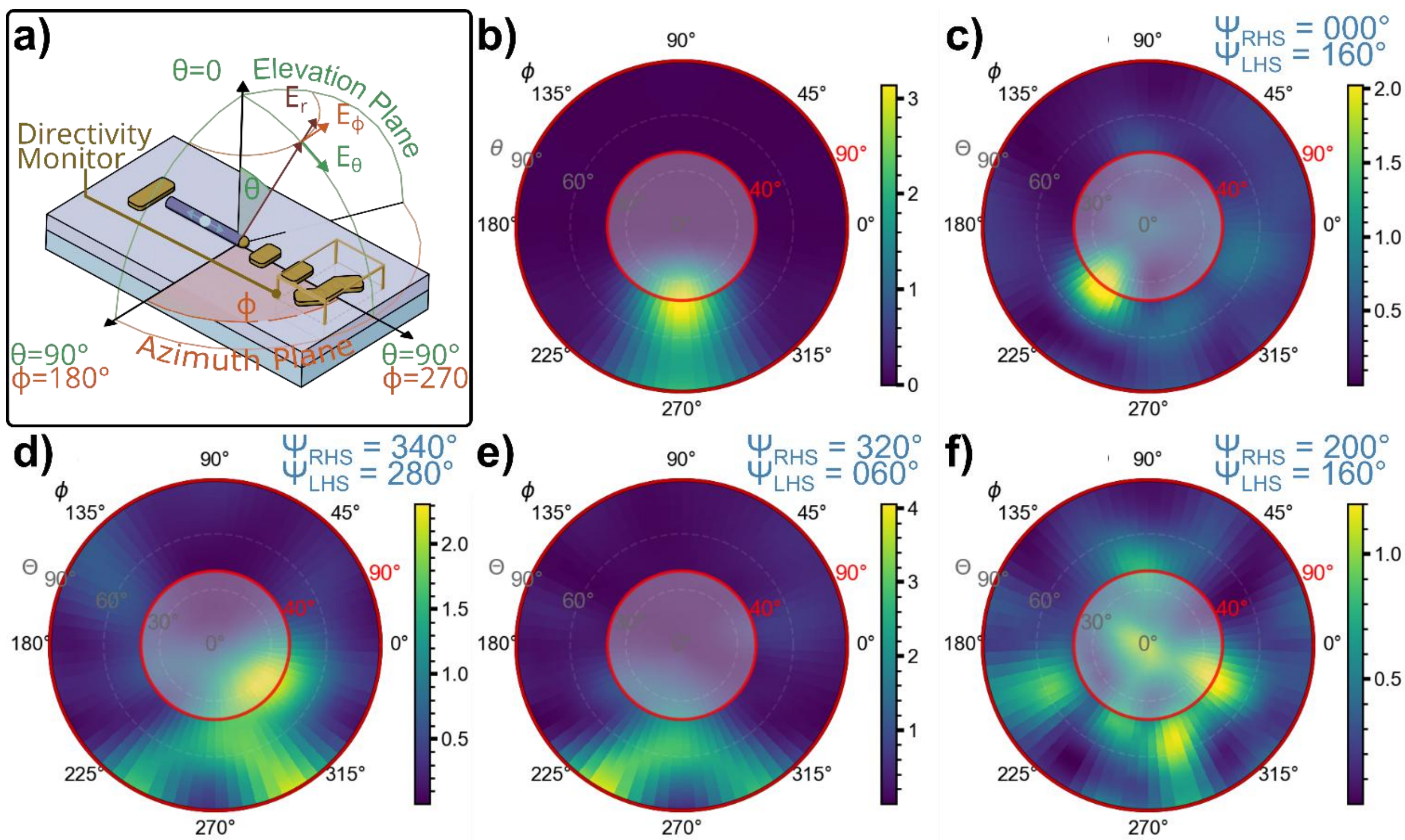

**Figure 3** –a) Model of modified Yagi-Uda antenna with asymmetric V-element in the end. Angles for the far-field angular emitter are given. b)-f) Angular plots of the total far-field radiated power density in each direction normalized by the total radiated power in all directions. $\phi$ angle from 0-360 is around the circumference. $\theta$ from 0-90 is given radially with 0 in the center and 90 at the edge. The emission in the angular range $\theta$ from 40-90 is indicated by the two red lines will be realistically available to the receiver (when the system is covered in a planar 2D waveguiding material (e.g. $Al_2O_3$). b) Is from an antenna less system, while in c)-f) Angles $\Psi_{RHS}$ and $\Psi_{LHS}$ refers to the orientation the wings of the V-element as given in Fig 1b. A wide range of forward emission patterns are possible depending on the asymmetry of the outermost element of the modified Yagi-Uda (see Fig 1 for notation).

We now go on to evaluate the effect of angular emission into the far-field from an asymmetric Yagi-Uda as shown in Fig 3. Directivity variation between V-element configurations is measured using a far-field monitor designed to determine the angular distributed far-field intensity (as seen in Fig. 3a). It projects far-field components $E_r, E_\theta$ and $E_\phi$ onto a sphere and relates them to the angular distribution of power radiated per unit solid angle $U(\theta,\phi)$. This is then normalized with the total radiated power $P_{rad}$ /4p. The angular plots of Fig 3b-f have the $\theta$ angle radially and the $\phi$ angle going around the circumference. The critical angles at the wavelengths of the light in the case of a quasi 2D ($Al_2O_3$) waveguiding film surrounding the full emitter-receiver is ~40 degrees. Thus, a simple estimate of the angular span of the light that can be transmitted to a far-field receiver would be 40-90 degrees which is indicated by the two red lines in the Fig 3b-f.

From these plots it is seen that the shape of the outermost element of the Yagi-Uda antenna can significantly affect the angular distribution of the radiated power. The distribution in $\phi$ shows that directions of the light away from the direct forward of the antenna are possible (Fig 3c) as well as the splitting of the light in two, three and four beams (Fig 3d-f). Estimates based on experimental data and further simulations[33] indicate that communication using even a single nanowire up to at least 100 μm is possible if all efficiencies are optimized. Thus, using such antennas, light could be directed towards several additional waveguiding elements for further reception and onwards communication.

After exploring individual components of a network, we now move to simulating a full network of 7 emitter nodes and 7 receiver nodes as seen in Fig 1d and e. The separation between the emitter and receiver active nanowire components is set to 7 microns. In this compact configuration that is contained within a 7x14 micron area all nodes are naturally connected by light. We can thus achieve 49 weighted connections with no wiring between nodes; the weights will be determined by the antenna design. Additionally, the nanowires could also be placed in a more complex geometric pattern (see e.g. [31,32]) to further change the coupling strengths. The values found for optimal coupling of the individual components are used in simulations. Initially we simulated a network consisting of the symmetric Yagi-Uda antenna (Fig 1d). In Fig 4a we show the E-field intensity in the plane of the nanowire receivers resulting from a full simulation of the Yagi-Uda network. The part of the complete network shown in Fig 4a is indicated in Fig 4b. The full simulation including both emitter and receiver parts is shown in the SI. As can be seen when turning on emitter 4, light reaches both the directly opposing receiver nanowire as well as the surrounding ones. The field variations of the emitted light are highly complex and depend on the exact placement of antenna objects as well as the nanowires themselves. This again highlights that highly variable weights can be achieved by tailoring the position and orientation of nanowires and nanoplasmonic elements.

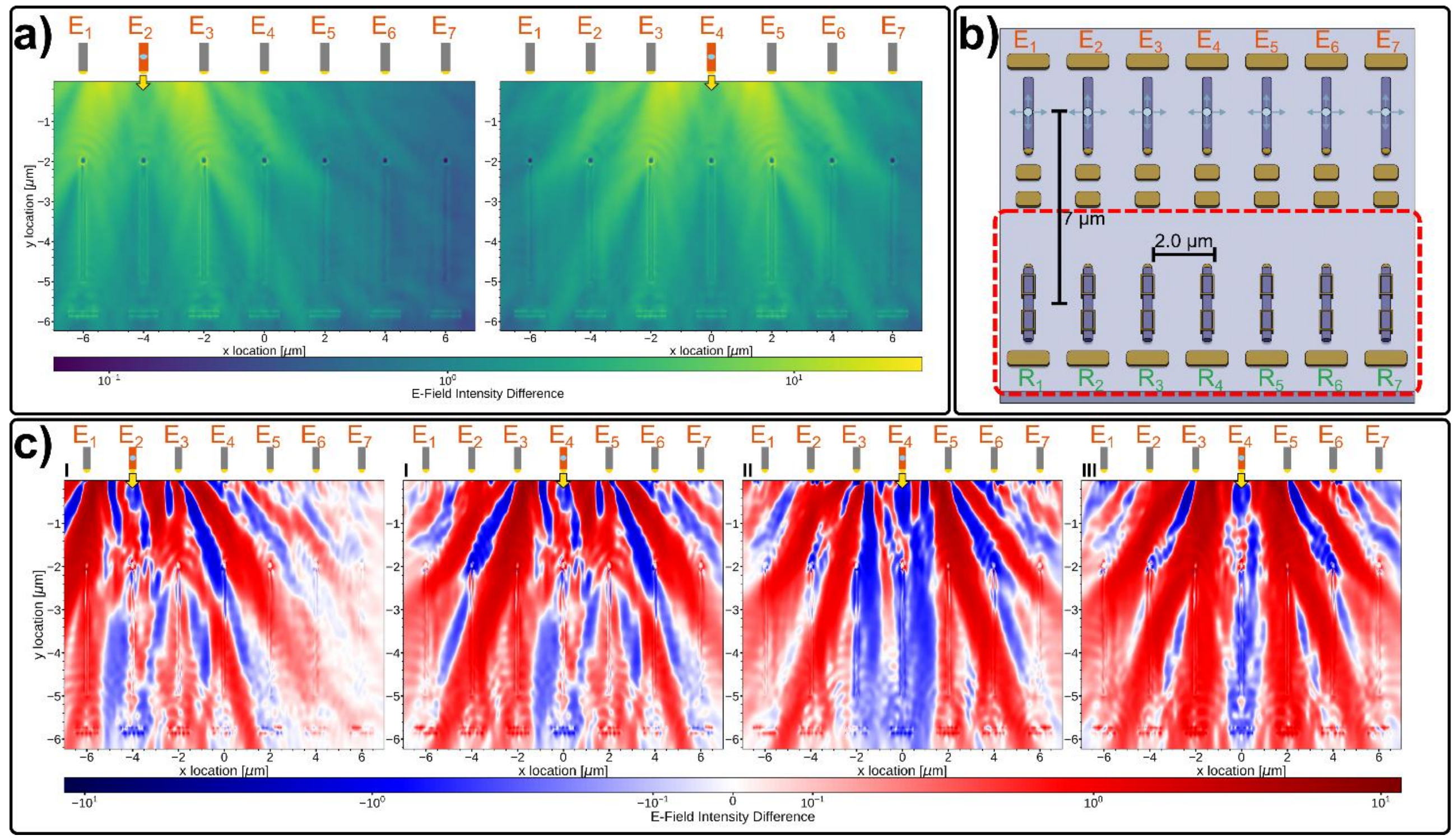

**Figure 4** - a) Intensity of the E-field in the receiver area of a system with 7 Yagi-Uda emitters and 7 receivers with examples of one emitter emitting light. The (red) emitting antenna is indicated by the yellow arrow (it is emitter #2 and #4). b) Sketch of the simulated emitter/receiver antenna system. The red box indicates the area of this system that is shown in Fig 4a. c) Intensity difference maps between symmetric Yagi-Uda and three modified Yagi-Uda systems with the middle emitter (#4) on. The design of I, II, III, referring to the wing angles of Fig 1b are shown in the SI and given by I:$\Psi_{RHS}$=[0,0,0,0,80,80,80], $\Psi_{LHS}$=[80,80,80,220,0,0,0];II:$\Psi_{RHS}$=[0,0,0,0,80,80,80],$\Psi_{LHS}$=[80,80,80,0,0,0,0]; III:$\Psi_{RHS}$=[0,0,0,0,80,30,80], $\Psi_{LHS}$=[30,70,80,320,60,30,0]. An angle 0 indicates that this wing is not present.

As nanowires have been explored as neural nodes in which different regions of the nanowire receive inhibiting and exciting signals[31,34] we can investigate how the Yagi-Uda antenna influences the connecting weights between nanophotonic neurons in such networks. To calculate communication weights for the nanowires [31,32] we assume a similar neural node design as in previous work[31,32]. In detail, it was shown that by introducing two photodiodes along a nanowire (which is experimentally feasible via p and n doping variation along the wire), regions in the nanowire that will give either an inhibiting or exciting contribution to a sigmoid nanowire neuron can be created[34]. Thus, the intensity in the two different regions of the nanowire will contribute either positive or negative weight connection between the first layer emitter nodes to the 2nd layer of receiver nodes. As a result, the weight matrix of the emitter connections to the receivers can be defined as:

$$\boldsymbol{W_{i,j}} = \frac{1}{n}\sum_{\boldsymbol{k=1}}^{\boldsymbol{n}} (\boldsymbol{Pabs_{activator,i,j,k}} - \boldsymbol{Pabs_{inhibitor,i,j,k}})$$

Power absorption in each segment is given by the two Pabs monitors as indicated in Figure 1d. Each emitter node $i$ is simulated separately, with resultant output node $j$ weights determined by the difference between activating and inhibiting power absorption. The emitter and receiver geometry is indicated in Fig 1d and e. The weight matric W can be used in simulations of neural networks based on the hardware components as in previous works[31,51] .

The total weight connection matrices for the 7-7 nanowire emitter receiver system for different antenna configurations are seen in Fig 5a. Individual values for activator and inhibitor regions are shown in the SI. With the present compact network, each emitter typically reaches the five closest nodes with significant weight assigned. The individual weights are small (less than a percent) as result of the design this can then allow light coupling to upwards a hundred nodes. To achieve a functioning network each neural node will then need an amplification >1000 from input to output. This is indeed an integral part of suggested nanowire based nanophotonic neurons[32] that operates with an intrinsic amplification of ~3000.

The distribution in connection weights is a consequence of both the emission from the individual wires as well as the interaction of the light with the assembly nanowires. The difference in the exciting and inhibiting matrix is a consequence of the highly varied field along the wire. The exact position of the emitter and receiver areas will then influence the weights since we are still in a region with near field effects. This agrees with previous work[51] that found that weights can be manipulated by translating and rotating the wires giving a baseline variability.

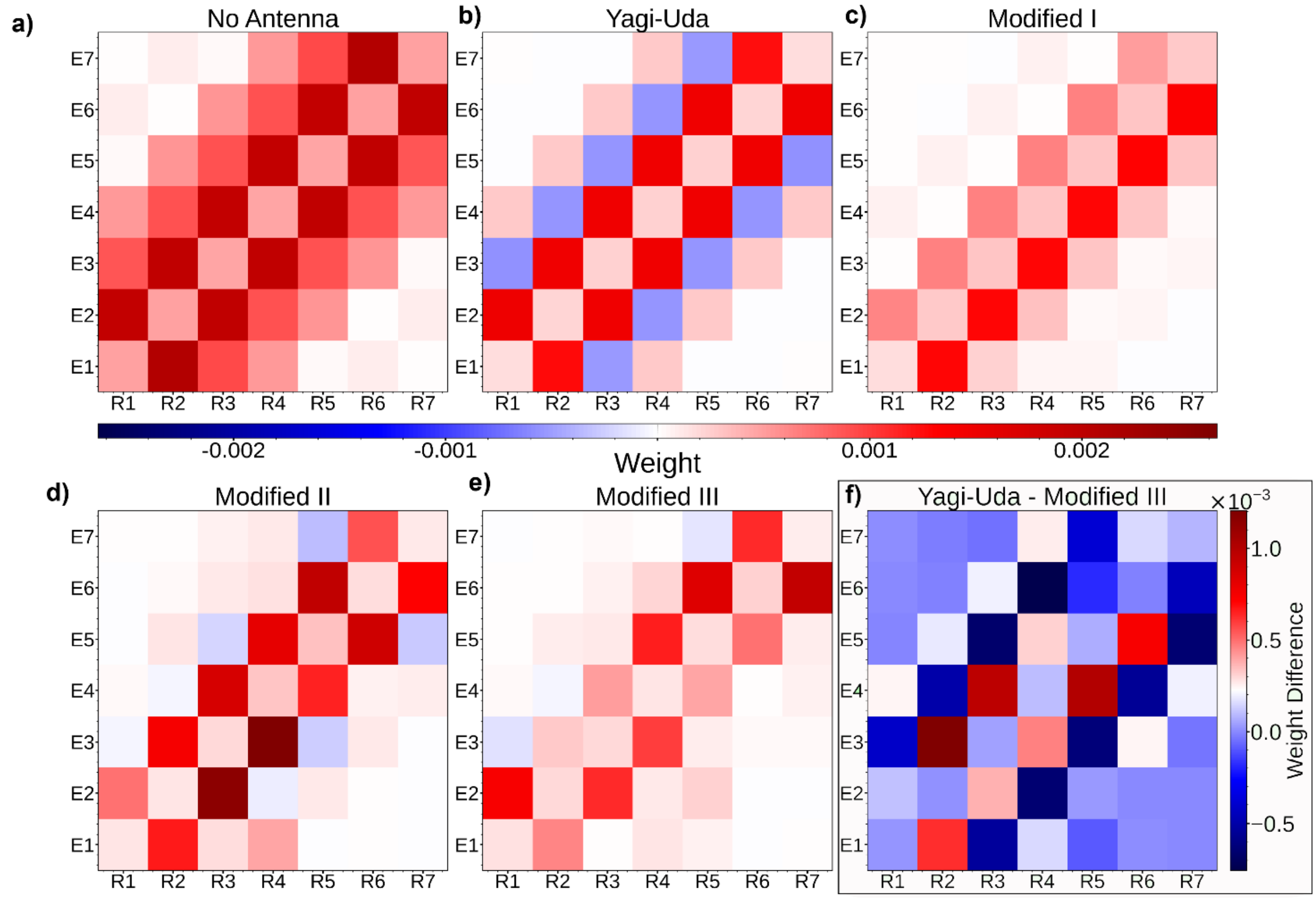

**Figure 5** - a) Weight connections of Emitters 1-7 on receivers 1-7 in networks as described by Fig 1d and e. The receiver nanowire is assumed to have both an inhibiting and exciting region as indicated in Fig 1d. The weights as defined by the summation of exciting and inhibiting signals are seen in a)-e). Schematics of the design of I, II, III, referring to Fig 1a are shown in the SI and given by I:$\Psi_{RHS}$=[0,0,0,0,80,80,80],$\Psi_{LHS}$=[80,80,80,220,0,0,0]; II:$\Psi_{RHS}$=[0,0,0,0,80,80,80],$\Psi_{LHS}$=[80,80,80,0,0,0,0]; III:$\Psi_{RHS}$=[0,0,0,0,80,30,80],$\Psi_{LHS}$=[30,70,80,320,60,30,0]. An angle 0 indicates that this wing is not present. f) Matrices showing the difference in weights between symmetric Yagi-Uda (Fig 5b) and the III asymmetric Yagi-Uda (Fig 5e).

Adding the Yagi-Uda antenna significantly changes the weight pattern, resulting in a focusing of the weights in the forward direction and even introducing negative weights for some wire couplings as seen inf Fig 5b. Then implementing the asymmetric Yagi-Udas, the weight matrices can be further manipulated creating even more complex coupling between the emitters and receivers as seen in Fig 5c-e. The introduction of asymmetry in the geometric configuration thus enables the creation of asymmetric weight matrices. To illustrate this clearer we also show the difference in weights between the symmetric and an asymmetric Yagi-Uda configuration in Fig 5f. The Yagi-Uda antenna thus adds additional means to control the weight distribution between nanowire based neural nodes when communicating in a frees pace configuration.

Going into more detail in Fig 4c we show the difference maps between the three asymmetric configurations (I,II,III of Fig 5) compared to the symmetric Yagi-Uda of Fig 4a indicating the significant differences in E-field for different shape of the Yagi-Uda V-elements. Configurations I and II have an identical asymmetry Yagi-Uda 1,2,3,5,6,7, while for I the emitter 4 is a standard Yagi-Uda while in II it is also asymmetric. As can be seen, even if emitter 4 is a symmetric Yagi-Uda (as in I), emission still changes significantly due to the changes in the surrounding antennas. For a compact network as the present this is a reasonable behavior as the neighboring antenna will influence each other. This indicates that to get a specific field distribution in the network iterative optimization of components will be necessary. Configuration III indeed represents a network in which Yagi-Uda asymmetries varies for the 7 emitters indicating in Fig 4c a more complex field variation. From the changes in weight patterns this indicates that weights can be significantly altered up to hundred percent and even switching from inhibiting to exciting or vice versa is possible.

While the E-field distributions are complex we can turn to the resulting calculated weights which in the end will be used in a network. These are shown in the three left columns of Fig 5a for three different configurations of V-elements across the 7 emitters (as denoted in the figure caption). From the weight matrices it is seen that a significant variety in weights can be achieved both for the exciting and inhibiting signals as well as the combined weights. To further illustrate this, we compare the asymmetric weight to the symmetric Yagi-Uda weights in Fig 5b showing the resulting difference weight matrices. Even this limited exploration of asymmetric elements shows that significant variation in weights can be achieved using an asymmetric element. Again, it is seen that changes in some emitter's antenna also affect other non-changed emitters. In the present situation, it is the magnitude of the weights that are mostly changed and not the sign of the weights (exciting/inhibiting). To change the sign the geometric placement of the inhibiting/exciting parts in the active nanowire elements could be used. Also as is expected, a focusing effect of the antenna is generally observed, but a substantial asymmetry across elements can be obtained.

## Conclusions

We have shown that Yagi-Uda antennas and modified asymmetric Yagi-Uda antenna of active III-V nanowires and passive Au islands can focus light in nanophotonic emitter-receiver communication systems. This can be used to focus communication between antenna nodes, but also to direct and split the light from a single emitter. Asymmetric nodes can be used to direct light towards different nodes in the far field. Simulating a condensed network with two layers of 7 nodes indicates that complex connectivity between nodes can be achieved by altering the shape of one of the Yagi-Uda director components. This can create compact small light communicating networks with variable weights. As the antenna does forward focusing, network connectivity will be strongest within a local cluster of nodes in the forward-facing direction. Such compact local networks are highly useful [31,32]: For compact networks with near-field effects, network design to achieve a certain weight pattern would be an iterative process in which antenna shapes and dimensions would be varied as well as the exact receiver position. Here also any electrical connectors to the wires can be considered and even actively used as scattering elements. By using antenna elements of both emitter and receiver as well as positioning highly complex network connectivity can be created. Thus, the design of an optical network can be used to hardwire weights of highly varying character into a network as is used in a e.g. navigation networks[52]. Introducing variable weights is possible introducing for example chemical elements that changes the plasmonic properties of the focusing elements[53] or introduction of all optical synaptic effects using dyes[54]. While the present work was done on specific active InP III-V nanowires in mind, the concept should be generally applicable to systems with nanoscale emitter/receiver optoelectronics.

## Author Contributions

V.F performed all FDTD simulations, data analysis, figure and diagram creation. V.F., and A.M. wrote the manuscript.

**Funding**

This work was supported by the Swedish Research Council, NanoLund, the Office of Naval Research (Grant No. N62909-20-1-2038) and the European Union Horizon Europe project InsectNeuroNano (Grant 101046790). Views and opinions expressed are, however, those of the author(s) only and do not necessarily reflect those of the European Union or the European Innovation Council. Neither the European Union nor the European Innovation Council can be held responsible for them.The authors declare no competing financial interests.